\documentclass[11pt, a4, reqno]{amsart}
\usepackage{parskip}
\usepackage{amsmath,amsthm,amssymb} 
\usepackage{graphics}
\usepackage{caption}
\usepackage{subcaption}
\usepackage{enumitem}
\usepackage{makecell}
\usepackage{color}
\usepackage[table]{xcolor}
\usepackage[calc]{picture}
\usepackage{blkarray}
\usepackage[english]{babel}
\usepackage{multirow}
\usepackage{diagbox}
\usepackage{xr}
\usepackage{lscape}
\usepackage[margin=1in]{geometry}
\usepackage{secdot}
\usepackage{rotating}
\usepackage{float}
\usepackage{setspace}
\usepackage{titlesec}
\usepackage{commath}
\usepackage[normalem]{ulem}
\usepackage[ruled,vlined]{algorithm2e}
\usepackage{amsfonts}
\usepackage{algcompatible}

\DeclareMathOperator*{\Prob}{P}

\DeclareMathOperator*{\logit}{logit}
\DeclareMathOperator*{\expit}{expit}
\DeclareMathOperator*{\bL}{\textbf{L}}
\DeclareMathOperator*{\bl}{\textit{\textbf{l}}}

\newcolumntype{?}{!{\vrule width 1pt}}

\newtheorem{theorem}{Theorem}

\newtheorem{example}{Example}
\newtheorem{corollary}{Corollary}[theorem]

\titlespacing*{\section}
{0pt}{1\baselineskip}{1\baselineskip}
\titlespacing*{\subsection}
{0pt}{1\baselineskip}{1\baselineskip}

\newcommand\independent{\protect\mathpalette{\protect\independenT}{\perp}}
\def\independenT#1#2{\mathrel{\rlap{$#1#2$}\mkern2mu{#1#2}}}

\begin{document}

\title[Stochastic treatment interventions in causal inference studies]{Intervention treatment distributions that depend on the observed treatment process and model double robustness in  causal survival analysis}
\author[Lan Wen, Julia L. Marcus and Jessica G. Young]{Lan Wen$^{\text{a}}$, Julia L. Marcus$^{\text{b}}$ \and Jessica G. Young$^{\text{b}}$ \vspace{0.6cm}
\\ \small
$^\text{a}$Department of Statistics and Actuarial Science, University of Waterloo\\ 200 University Ave W, Waterloo, ON N2L 3G1 \vspace{0.3cm} \\
$^\text{b}$Department of Population Medicine, Harvard Medical School, \\Boston, Massachusetts, 02215, U.S.A.\\ 
}
\date{}

\maketitle

\begin{abstract}
The generalized g-formula can be used to estimate the probability of survival under a sustained treatment strategy. When treatment strategies are deterministic, estimators derived from the so-called efficient influence function (EIF) for the g-formula will be doubly robust to model misspecification. In recent years, several practical applications have motivated estimation of the g-formula under non-deterministic treatment strategies where treatment assignment at each time point depends on the observed treatment process. In this case, EIF-based estimators may or may not be doubly robust. In this paper, we provide sufficient conditions to ensure existence of doubly robust estimators for intervention treatment distributions that depend on the observed treatment process for point treatment interventions, and give a class of intervention treatment distributions dependent on the observed treatment process that guarantee model doubly and multiply robust estimators in longitudinal settings. Motivated by an application to pre-exposure prophylaxis (PrEP) initiation studies, we propose a new treatment intervention dependent on the observed treatment process. We show there exist 1) estimators that are doubly and multiply robust to model misspecification, and 2) estimators that when used with machine learning algorithms can attain fast convergence rates for our proposed intervention. Theoretical results are confirmed via simulation studies.\\
{\textbf{keywords}}: {{Causal inference}, Double robustness, Estimating equations, Observational study, Stochastic treatment strategies}
\end{abstract}


\section{Introduction}

The goal of many observational analyses is to estimate the causal effect on survival of different time-fixed or time-varying treatment strategies, interventions or rules in a study population.  These causal effects can be formally defined by a contrast (e.g., difference or ratio) in the distributions of counterfactual outcomes had  interventions been implemented to ensure those strategies are followed in that population. Robins (1986) \cite{Robins1986} showed that, under assumptions that allow complex longitudinal data structures such that measured time-varying confounders may themselves be affected by past treatment, the \textsl{g-formula} indexed by a particular treatment strategy identifies the average counterfactual outcome under that strategy.  Therefore, estimators of the g-formula and associated contrasts indexed by different strategies may be used to estimate causal effects.

In practice, the g-formula typically depends on high-dimensional nuisance parameters.  In this case, many estimators of the g-formula and associated contrasts have been proposed including the density-based parametric g-formula \cite{Robins1986}, iterated conditional expectation (ICE) estimators \cite{Tran2019,Wen2021}, inverse probability weighted (IPW) estimators \cite{Cain2010, Neugebauer2014}, and estimators derived from the efficient influence function (EIF) \cite{Bang2005, van2011}. EIF based estimators (i.e., estimators constructed to evaluate the EIF from an empirical sample) have several theoretical advantages over the other approaches including they may be $\sqrt{n}$-consistent if the nuisance functions are estimated at slower rates through flexible nonparametric or machine learning methods \cite{Robins2009Q,Robins2016,Chernozhukov2018}.  

EIF based estimators may also have a model double-robustness property in that, when nuisance functions are estimated via {parametric} models, these estimators may remain consistent and asymptotically normal if models for only one of two (sets of) nuisance functions are correctly specified, not necessarily both. This model double robustness property always holds for EIF estimators when the g-formula is indexed by a \textsl{deterministic treatment strategy} at most dependent on past treatment and confounders measured in the observational study \cite{Bang2005, Stitelman2012, Rotnitzky2017}.  However, the identification results of Robins (1986)\cite{Robins1986} were not limited to such deterministic  strategies but generalized to allow identification of \textsl{stochastic} treatment strategies at most dependent on this measured past.  The latter identifying functional or \textsl{generalized g-formula} depends on the \textsl{intervention treatment distribution}, that is, the distribution of treatment under an intervention that ensures the strategy of interest is followed conditional only on the measured past in the observational study.  The generalized g-formula coincides with the more familiar g-formula indexed by a deterministic strategy when the intervention treatment distribution is chosen as degenerate conditional on any level of the measured past.

Recently, several practical applications have motivated estimation of the generalized g-formula indexed by intervention treatment distributions that depend on the \textsl{observed treatment process}, that is, the observed treatment distribution conditional on the measured past \cite{ Taubman2009, Munoz2012, Haneuse2013, Kennedy2019, Young2019}. The generalized g-formula indexed by an intervention treatment distribution dependent on the observed treatment process has the particular advantage of relying on relatively weak positivity conditions \cite{Haneuse2013,Young2014} even, for example, in observational studies where the propensity score is equal or close to zero for certain measured confounder histories \cite{Kennedy2019}. When the (degenerate or non-degenerate) intervention treatment distribution does \textsl{not} depend on the observed treatment process,  EIF derived estimators of the generalized g-formula will be model doubly robust.  However, when the intervention treatment distribution \textsl{does} depend on the observed treatment process, such estimators may or may not be doubly robust \cite{Munoz2012, Haneuse2013,Kennedy2019,Diaz2020}.

In this paper, we exploit particular representations of the generalized g-formula to give sufficient conditions for the existence of doubly robust estimators for point treatment interventions when the chosen intervention treatment distribution depends on the observed treatment process with examples from the recent literature. We also provide a general form of EIFs for a class of intervention treatment distributions that may depend on the observed treatment process in longitudinal settings that guarantee model doubly (and multiply) robust estimators. Motivated by observational studies of the effects of realistic HIV pre-exposure prophylaxis (PrEP) initiation interventions, we consider a new class of intervention treatment distributions dependent on the observed treatment process that is a variation on the incremental propensity score interventions proposed by Kennedy (2019)\cite{Kennedy2019}. We show that estimators based on the EIF for our proposed  intervention treatment distribution are model doubly/multiply robust, and can attain fast convergence rates even when used in combination with machine learning algorithms, where modelling assumptions are relaxed. We illustrate both EIF-based, as well as simpler singly robust, estimators of the g-formula indexed by this class of intervention treatment distribution in simulated data and in an illustrative data application. 

\section{Observed data structure}
\label{sec:observedata}
Consider a longitudinal study with $j=0,1,2,\ldots, J$ denoting a follow-up interval (e.g., week, month) where $J$ is the end of the follow-up of interest. Assume the following random variables are measured in this study on each of $n$ individuals meeting some eligibility criteria at baseline.  For each $j=0,1,2,\ldots, J-1$, let $A_j$ denote binary or discrete treatment variable during interval $j$, $\bL_j$ a vector of additional time-varying covariates measured in interval $j$, and $Y_{j+1}$ an indicator of survival by interval $j+1$. 
For notational simplicity, we will assume throughout that all covariates are discrete in that they have distributions that are absolutely continuous with respect to a counting measure but arguments naturally extend to settings with continuous covariates and Lebesgue measures.
By definition,  ${Y}_0 = 1$ (all individuals are at risk of failure baseline) and by convention we define $\bar{\bL}_{-1} = \bar{A}_{-1} = \emptyset$. For a random variable $X$, we let $\bar{X}_j = (X_0,\ldots, X_j)$ denote history through time $j$. We assume the ordering $O=(\bL_0,A_0, Y_1,\ldots, \bL_{J-1}, A_{J-1}, Y_J)$.  Without loss of generality, we will assume no individual is lost to follow-up until Section \ref{sec:censoring}.

\section{Intervention treatment distribution}
\label{sec:assumptionsetc}
Let $g$ denote a treatment rule that specifies how treatment should be assigned at each $j=0,1,2,\ldots, J$. Following Richardson and Robins (2013) \cite{Richardson2013}, denote $(\bL^{g}_j,Y^{g}_j)$ and $A_j^{+g}$ as the natural values of covariates and survival status and the intervention value of treatment at $j$ under $g$, respectively.  In turn, the distribution of $A^{+g}_j$ evaluated at some treatment level $a_j$ conditional on the ``measured past'' under $g$ $(Y^{g}_{j}=1,\overline{\bL}^{g}_j=\overline{\bl}_j,\overline{A}^{+g}_{j-1}=\overline{a}_{j-1})$ is specified by 
$q^{g}(a_j\mid 1, \bar{\bl}_{j}, \bar{a}_{j-1})\equiv \Pr(A^{+g}_j=a_j\mid Y^{g}_{j}=1,\overline{\bL}^{g}_j=\overline{\bl}_j,\overline{A}^{+g}_{j-1}=\overline{a}_{j-1})$
which we refer to as the \textsl{intervention treatment distribution} at $j$ associated with $g$. 

When treatment assignment at any time under a selected rule $g$ deterministically depends on the measured past, there is only one value $a^{+}_j\in \text{supp}(A^{+g}_j)$ given any history $(\bar{\bl}_{j}, \bar{a}_{j-1})\in \text{supp}(\overline{\bL}^{g}_j,\overline{A}^{+g}_{j-1})$ for those with $Y^{g}_j=1$.  In this case, $q^{g}(a_j\mid 1, \bar{\bl}_{j}, \bar{a}_{j-1})=1$ when $a_j=a^{+}_j$ and 0 otherwise. Examples include \textsl{static} deterministic rules that assign the same level of treatment to all surviving individuals at all follow-up times 
and \textsl{dynamic} deterministic rules that assign treatment based on the measured past. 

By contrast, when a selected rule $g$ assigns treatment stochastically at some $j$ (as a random draw from a distribution), at most dependent on the measured past, then there will be multiple values $a^{+}_j\in \text{supp}(A^{+g}_j)$ such that we may have $0<q^{g}(a_j\mid 1, \bar{\bl}_{j}, \bar{a}_{j-1})<1$ when $a_j=a^{+}_j$.  We focus here on the problem of estimating $\mbox{E}[Y_J^{g}]=\Pr[Y_J^{g}=1]$, the cumulative survival probability by end of follow-up under a choice of $g$, when the intervention treatment distribution associated with $g$ has this non-degenerate property of a stochastic rule, in particular, \textsl{through its dependence on the observed treatment distribution conditional on the measured past} as specified by $
f(a_j\mid 1, \bar{\bl}_{j}, \bar{a}_{j-1})\equiv \Pr(A_j=a_j|Y_{j}=1,\overline{\bL}_j=\overline{\bl}_j,\overline{A}_{j-1}=\overline{a}_{j-1}).$ 
We refer to $f(a_j\mid 1, \bar{\bl}_{j}, \bar{a}_{j-1})$ as the \textsl{observed treatment process} conditional on the measured past.  The observed treatment process evaluated at $a_j=1$ coincides with the so-called \textsl{propensity score} \cite{Rosenbaum1983} at $j$ when treatment $A_j$ is binary. Next, we consider several examples of such intervention distributions. 

\section{Examples of intervention treatment distributions that depend on the observed treatment process}
\label{examples}
In this section, we review examples of intervention distributions considered in the literature that have depended on the observed treatment process:
\begin{itemize}
\item \textsl{Dynamic treatment initiation strategies with grace period:} 

Motivated by questions about the effects of CD4-based treatment initiation strategies, previous authors have considered strategies of the form ``If a condition for treatment initiation is met by interval $j$ then start treatment by $m+j$ for a selected grace period $m$, with no intervention in intervals $j$ through $j+m-1$. Otherwise, do not start at $j$'', $\forall j$ \cite{Cain2010,Young2011}. 
For $A_j$ an indicator of treatment initiation by $j$ and $L_j^\ast\in \bL_j$ an indicator that the condition for initiating treatment has been met by $j$, the intervention treatment distribution at each $j$ is 
\[
   q^{g}(1\mid 1, \bar{\bl}_{j}, \bar{a}_{j-1}) \left.
  \begin{cases}
    1, & \text{if } l^\ast_{j-m}=1\\
    0, & \text{if } l^\ast_{j}=0\\
    f(1\mid 1, \bar{\bl}_{j}, \bar{a}_{j-1}), & \text{otherwise}
  \end{cases}
  \right.
\]
or
$q^{g}(a_j\mid 1, \bar{\bl}_j, \bar{a}_{j-1}) = (1-l_j^\ast)(1-a_j) + l_{j-m}^\ast a_j + (1-l_{j-m}^\ast)l_j^\ast f(a_j\mid 1, \bar{\bl}_{j}, \bar{a}_{j-1})$.

\item \textsl{Representative interventions:}

Motivated by observational studies to understand the long-term effects of lifestyle interventions (e.g., interventions that increase daily minutes of physical activity), previous authors have considered \textsl{representative interventions} that assign the value of a multi-level treatment to an individual at each $j=0,\ldots,J$ as a random draw from a particular distribution:  specifically, the observed distribution of treatment in interval $j$ among those who, in the  observational study, (i) had the same measured confounder and treatment history prior to $j$ as that individual and (ii) had treatment at $j$ at or above a cutoff $\delta$ (or more generally, within a pre-specified range), e.g., ``at least 30 minutes of daily physical activity'' \cite{Picciotto2012,Young2019}. In this case, 
\[
q^{g}(a_j\mid 1, \bar{\bl}_j, \bar{a}_{j-1}) = f(a_j\mid 1, \bar{\bl}_j, \bar{a}_{j-1},a_j\geq \delta)
\]
This intervention distribution notably only depends on the observed treatment process at $j$ \textsl{among those with treatment in the pre-specified range at $j$}.

\item \textsl{Deterministic interventions that depend on the natural value of treatment:}

Alternative interventions that maintain a multi-level treatment within a pre-specified range have been posed that assign treatment at each $j$ as a function of the natural treatment value at $j$ \cite{Robins2004,Taubman2009}, e.g., ``If the natural value of treatment at $j$ is below $\delta$ than intervene and set treatment at $j$ to $\delta$.  Otherwise, do not intervene at $j$''.  The resulting intervention distribution at each $j$ (conditional only on the measured past and marginal with respect to the natural value of treatment at $j$) is
\begin{equation*}
q^{g}(a_j\mid 1, \bar{\bl}_{j}, \bar{a}_{j-1})=F_{A_j}(\delta \mid 1, \bar{\bl}_{j}, \bar{a}_{j-1}) I(a_j=\delta) + I(a_j\geq \delta) f(a_j\mid 1, \bar{\bl}_{j}, \bar{a}_{j-1})
\end{equation*}
where $F_{A_j}(\delta \mid 1, \bar{\bl}_{j}, \bar{a}_{j-1})=\sum_{a_j<\delta}f(a_j\mid 1, \bar{\bl}_{j}, \bar{a}_{j-1})$. 

\item \textsl{Incremental propensity score interventions: }
Kennedy (2019)\cite{Kennedy2019} posed \textsl{incremental propensity score interventions} that at each $j$ assign a binary treatment according to a strategy $g$ that results in an intervention treatment distribution defined by a shifted version (on the odds scale) of the propensity score.  Specifically, for a particular $\delta\in (0,\infty)$
\begin{equation}
   q^{g}(1\mid 1, \bar{\bl}_{j}, \bar{a}_{j-1}) = 
    \dfrac{\delta f(1\mid 1, \bar{\bl}_{j}, \bar{a}_{j-1})}{\delta f(1\mid 1, \bar{\bl}_{j}, \bar{a}_{j-1}) + f(0\mid 1, \bar{\bl}_{j}, \bar{a}_{j-1})}
  \label{kennedyshift}
\end{equation}
or $q^{g}(a_j\mid 1, \bar{\bl}_j, \bar{a}_{j-1}) = \{{a_j \delta f(a_j\mid 1, \bar{\bl}_{j}, \bar{a}_{j-1}) + (1-a_j)f(a_j\mid 1, \bar{\bl}_{j}, \bar{a}_{j-1})}\}\{\delta f(a_j\mid 1, \bar{\bl}_{j}, \bar{a}_{j-1}) + f(1-a_j\mid 1, \bar{\bl}_{j}, \bar{a}_{j-1})\}^{-1}$.
Here we consider a modification of (\ref{kennedyshift}) motivated by the PrEP context.  Specifically, for $\delta\in [0,1]$ and $L_j^\ast\in \bL_j$, a measured marker of risk for HIV acquisition (e.g., receiving a bacterial sexually transmitted infections or STI test at $j$ and no prior HIV diagnosis), we consider interventions indexed by the alternative intervention treatment distribution
\begin{equation}
   q^{g}(0\mid 1, \bar{\bl}_{j}, \bar{a}_{j-1}) = \left.
  \begin{cases}
    f(0\mid 1, \bar{\bl}_{j}, \bar{a}_{j-1}), & \text{if } l_j^\ast=0 \\
    f(0\mid 1, \bar{\bl}_{j}, \bar{a}_{j-1})\delta, &  \text{if } l^\ast_{j}=1
 \end{cases}
  \right.
  \label{ourshift}
\end{equation}
or $q^{g}(a_j\mid 1, \bar{\bl}_j, \bar{a}_{j-1}) = (1-\delta) l_j^\ast a_j  + (l_j^\ast \delta + 1 - l_j^\ast) f(a_j\mid 1, \bar{\bl}_{j},\bar{a}_{j-1})$ after some algebra.

In words, the probability of initiating treatment conditional on the measured past under $g$ at each $j$ will be larger than the propensity score at $j$ (that under no intervention) by decreasing its complement by a factor of $\delta$ for those with an indication ($L_j^\ast=1$).
Specifically, $\delta$ is the risk ratio for \textsl{not initiating} treatment under $g$ vs. no intervention conditional on the measured past. Choosing $\delta=0$ corresponds to ``always treat" those with $L_j^\ast=1$, and $\delta=1$ to no intervention.  We will refer to interventions indexed by either (\ref{kennedyshift}) or (\ref{ourshift}) as  \textsl{incremental propensity score interventions}, distinguishing them by the classifier \textsl{odds shift} or \textsl{multiplicative shift}, respectively.
\end{itemize}

\section{Identification by the generalized g-formula}
\label{sec:g-formula}
Consider a treatment assignment rule $g$ at most dependent on the measured past.  Further, let $\mathcal{D}_g$ denote the set of all \textsl{deterministic} strategies at most dependent on this past that individuals could be observed to follow under the selected rule $g$, with $d$ any element of $\mathcal{D}_g$.  In the special case when $g$ is initially selected to be a deterministic rule then the only element of $\mathcal{D}_g$ is $g$.  Otherwise, $\mathcal{D}_g$ may contain many elements.  
Let $Y_{j}^{d},L_j^{d}$ and $A_j^{+_d}$ denote the natural values of survival status and covariates and the intervention value of treatment at $j$, respectively, under a  deterministic $d\in\mathcal{D}_g$ though ($j=0,\ldots, J)$ and consider the following assumptions:
\begin{enumerate}
\item Exchangeability: 
$(Y_{j+1}^{d}, \ldots, Y_{J}^{d})\independent A_j \mid \bar{\bL}_j=\bar{\bl}_j, \bar{A}_{j-1}=\bar{a}_{j-1}^{+}, Y_j=1$.
\item Consistency: If $\bar{A}_{j} = \bar{A}_j^{+_d}$ then $\bar{Y}_{j+1} = \bar{Y}_{j+1}^{d}$ and $\bar{\bL}_{j} = \bar{\bL}_{j}^{d}$
\item Positivity: $f_{\bar{\bL}_j,\bar{A}_{j-1},Y_j} (\bar{\bl}_j, \bar{a}_{j-1}^{+},1) >0  \Longrightarrow f_{A_j\mid Y_j, \bar{\bL}_j,  \bar{A}_{j-1}} (a_j^{+} \mid 1, \bar{\bl}_j,\bar{a}_{j-1}^{+})>0$
\end{enumerate}
Robins (1986) \cite{Robins1986} showed that given these exchangeability, consistency and positivity conditions hold for all deterministic $d\in\mathcal{D}_g$ then the following function of only the observed data identifies $\mbox{E}[Y_J^{g}]$:
\begin{align}
\psi^{g}=&\sum_{\forall\bar{a}_{J-1}} \sum_{\forall\bar{\bl}_{J-1}} \Prob(Y_{J}=1\mid {Y}_{J-1} = 1, \bar{\bL}_{J-1}=\bar{\bl}_{J-1}, \bar{A}_{J-1}=\bar{a}_{J-1})
\times  \label{eq:gform0} \\[-0.5em]
&\prod_{s=0}^{J-1} \Prob(Y_s=1\mid {Y}_{s-1} =1, \bar{\bL}_{s-1}=\bar{\bl}_{s-1}, \bar{A}_{s-1}=\bar{a}_{s-1})f({\bl}_s\mid {Y}_{s} = 1, \bar{\bl}_{s-1}, \bar{a}_{s-1})q^{\text{\text{g}}}(a_s\mid  1, \bar{\bl}_{s}, \bar{a}_{s-1}) \nonumber
\end{align}
The function $\psi^{g}$ is referred to as the \textsl{generalized g-formula} indexed by the intervention treatment distribution $q^{\text{\text{g}}}(a_s\mid 1, \bar{\bl}_{s}, \bar{a}_{s-1})$.  Note that, under stronger identifying conditions, the generalized g-formula may identify the outcome mean under a treatment rule $g$ that depends on more than the measured past \cite{Richardson2013,Young2014}.  Also see Web Appendix B.

\subsection{Generalized positivity}
Note the assumption that the positivity condition above holds for all deterministic $d\in\mathcal{D}_g$ can be equivalently stated as follows:  
\begin{equation}
 q^{g}(a_j\mid 1, \bar{\bl}_{j}, \bar{a}_{j-1}) >0  \Longrightarrow f(a_j\mid 1, \bar{\bl}_{j}, \bar{a}_{j-1})>0   \label{genpositivity}
\end{equation}
for all $\overline{\bl}_j,\overline{a}_{j}\in \text{supp}(\overline{\bL}^{g}_j,\overline{A}^{+g}_{j})$.  
The positivity condition (\ref{genpositivity}) generalizes the more familiar definition of positivity often relied on in the literature that there may be treated and untreated individuals within any level of the measured past; i.e., the assumption that the propensity score and its complement are positive for all possible measured histories and all $j$.  It is straightforward to see that the more general condition \eqref{genpositivity}  reduces to this typical definition of positivity only for the special case of a static deterministic intervention $g$ on a binary treatment.  By contrast, the more general condition \eqref{genpositivity} only requires that, for any level of the past possible in the observational study and also plausible under $g$, if an intervention level of treatment can occur under $g$ it must also possibly occur in the observational study.  Depending on the choice of $g$, this condition may hold when traditional definitions requiring positive propensity scores fail.  Intervention treatment distributions that depend on the observed treatment process may help avoid positivity violations by this more general definition and, in some instances, may guarantee that positivity violations cannot occur regardless of the observed treatment process. We discuss this further in the next section.

Similar to arguments given in Kennedy (2019)\cite{Kennedy2019}, the odds shift (\ref{kennedyshift}) has the particular advantage that, by construction, the generalized positivity condition (\ref{genpositivity}) is guaranteed to hold, no matter the nature of the observed treatment process. By contrast, the multiplicative shift (\ref{ourshift}) only enjoys this guarantee for measured pasts consistent with $L_j^\ast=0$.  However, compared to (\ref{kennedyshift}) which is indexed by a shift $\delta$ with no upper bound that quantifies an odds ratio, (\ref{ourshift}) may be easier to communicate to subject matter collaborators as it constrains the choice of $\delta\in[0,1]$ and quantifies a risk ratio. Notably, the performance of weighted estimators of $\psi^{g}$ indexed by both (\ref{kennedyshift}) and (\ref{ourshift}) are relatively resilient to so-called ``near positivity violations'' -- such that (\ref{genpositivity}) holds but $f(a_j\mid 1, \bar{\bl}_{j}, \bar{a}_{j-1})$ is still close to zero for some $(\bar{\bl}_{j}, \bar{a}_{j-1})$ -- particularly when $\delta$ is chosen to coincide with relatively small increases in treatment uptake under $g$ (see Section \ref{sec:sim}). 
 
\section{Model double robustness when the intervention treatment distribution depends on the observed treatment process}
\label{sec:mainint}
Suppose that the observed data $O$ defined in Section \ref{sec:observedata} follows a law $P$ which is known to belong to $\mathcal{M}=\{P_\theta:\theta\in \Theta\}$ where $\Theta$ is the parameter space. The efficient influence function (EIF) $U_{\psi^g}(O)$ for the causal parameter $\psi^{g}\equiv \psi^g(\theta)$ in a non-parametric model that imposes no restrictions on the law of $O$ is given by
${d\psi^g(\theta_t)}/{dt}\vert_{t=0} = E\{U_{\psi^g}(O)S(O)\}$, where ${d\psi^g(\theta_t)}/{dt}\mid_{t=0}$ is known as the pathwise derivative of the parameter $\psi^g$ along a parametric submodel of the observed data distribution indexed by $t$, and $S(O)$ is the score function of the parametric submodel evaluated at $t=0$ \cite{newey1994,Van2000}.
In this section, we provide results that aid the intuition on the existence of doubly robust estimators of $\psi^{g}$ when the intervention treatment distribution depends on the observed treatment process through understanding properties of the EIF for the parameter $\psi^{g}$.  

\subsection{Point treatment}
We begin with the special case of a point treatment where $J=1$ and $O=(\bL_0,A_0,Y_1)\equiv(\bL,A,Y)$.  In this case, (\ref{eq:gform0}) reduces to
$\psi^g= \sum_{\forall\bar{a}} \sum_{\forall\bar{\bl}}E(Y|A=a,\bL=\bl)q^g(a|\bl)f(\bl)$.

\begin{theorem}
Suppose  $\psi^g$ can be written as a linear combination of the form:
\begin{equation}
\psi^g=c_{1}\underbrace{E\{h_{1}(O)\}}_{\nu_{1}} + c_{2} \underbrace{E[E\{h_{2}(O)\mid A=a^\ast,L \} ]}_{\nu_{2}}
\label{eq:thm1eq}
\end{equation}
where $a^\ast$, $c_1$ and $c_2$ are constants, and $h_{1}(O)$ and $h_{2}(O)$ are known measurable functions of $O$ (i.e., they do not depend on $\theta$). Then the EIF for $\psi^g$ is given by 
\begin{equation}
U_{\psi^g}(O) = c_{1}h_{1}(O) + c_{2} \left[\frac{I(A = a^\ast)}{f(A\mid \bL)} \left[h_{2}(O) - E\{h_{2}(O)\mid A,\bL\}\right] + E\{h_{2}(O)\mid A=a^\ast,\bL\} \right]-\psi^g
\label{eq:thm1eq2}
\end{equation}
\label{theorem:thmone}
\end{theorem}
\vspace{-2em}
See Web Appendix A for proof. 
Clearly, $\psi^{g}$ under a static deterministic strategy that sets treatment to level $a^\ast$ for all individuals trivially meets the conditions of Theorem 1  by selecting $h_1(O)=0$, $c_{2}=1$, $h_{2}(O)=Y$. In this case, the EIF for the g-formula indexed by $g$ or $E\{E(Y\mid A=a^\ast,\bL)\}$ equals:
\begin{equation}
U_{\psi^g}(O)=\frac{I(A = a^\ast)}{f(A\mid \bL)} \left\{Y - m(A,\bL)\right\} + m(a^\ast,\bL) - \psi^{a^\ast}
\label{eq:eifstatic}
\end{equation}
where $m(A,\bL) \equiv E(Y\mid A,\bL)$ and $m(a^\ast,\bL) \equiv E(Y\mid A=a^\ast,\bL)$\cite{Bang2005,Tsiatis2006,van2011}. 
A heuristic justistification for Theorem \ref{theorem:thmone} follows from the fact that the EIF of ${\nu_{1}}\equiv \nu_{1}(\theta)$ is simply $h_{1}(O)$, and the EIF of ${\nu_{2}}\equiv \nu_{2}(\theta)$ can realized by replacing $Y$ with $h_{2}(O)$ in Expression (\ref{eq:eifstatic}), \textit{as the function $h_{2}(\cdot)$ does not depend on $\theta$ and therefore its pathwise derivative is zero}.
Furthermore, it is established that an estimator derived from the influence function (\ref{eq:eifstatic}) (e.g., an estimator solving $\sum_{i=1}^{n}U_{\psi^g}(O_i)=0$ for $\psi^g$) is model doubly robust in that it remains consistent if estimated under correctly specified parametric models for either one of two (sets of) nuisance functions, specifically $E(Y\mid A,\bL)$ or $f(A\mid \bL)$. 
The following Corollary gives a sufficient condition for the existence of doubly robust estimators of  $\psi^g$ when the intervention treatment distribution depends on the observed treatment process, provided the conditions of Theorem 1 hold. 

\begin{corollary}
Suppose the conditions of Theorem \ref{theorem:thmone} hold.  If {$h_{2}(O) = Y \tilde{h}_{2}(A,\bL)$}, where  $\tilde{h}_2(A,\bL)$ is a known measurable function of $(A,\bL)$, then an estimator of $\psi^g$ derived from an EIF of the form (\ref{eq:thm1eq2}) is model doubly robust.
\label{corollary:corone}
\end{corollary}
A proof of Corollary \ref{corollary:corone} is given in Web Appendix A. 
A similar heuristic reasoning for Corollary \ref{corollary:corone} is that the estimator of the EIF of a mean outcome does not rely on any models, and doubly robust estimators exist for $\nu_{2}$ because we have simply replaced $Y$ with $h_{2}(O)$ in Equation (\ref{eq:eifstatic}) which does not depend on $\theta$.
We now consider some applications of Theorem 1 and Corollary 1.1 to examples where the intervention distribution indexing $\psi^g$ depends on the observed treatment process. 

\setcounter{example}{0}
\begin{example}
Consider a variation of the grace period treatment initiation strategies defined in Section \ref{examples} for $J=1,~m=0$, such that, rather than withholding treatment when $L^\ast=0$, no intervention is made. The intervention treatment distribution is then given by $q^g(a\mid \bl)=(1-l^\ast)f(a\mid \bl) + l^\ast a $ .
\end{example}
For this choice of intervention distribution we have:
\begin{align*}
\psi^g &= E_{\bL}\left\{\sum\nolimits_{a=0}^1 E(Y\mid a,\bL)q^g(a\mid \bL)\right\}
\\& = E_{\bL}\left[\sum\nolimits_{a=0}^1 E(Y\mid a,\bL)\left\{(1-L^\ast)f(a\mid \bL) + I_{\{1\}}(a)L^\ast  \right\}\right]  \nonumber
\\& = {E_{\bL,A}\left[E\left\{Y(1-L^\ast)\mid A,\bL \right\}\right]} + {E_{\bL}\left[E\left\{YL^\ast\mid A=1,\bL \right\}\right]}
\end{align*}
Selecting $a^\ast=1$, $c_{1}=c_{2}=1$, $h_{1}(O)=Y(1-L^\ast)$, and $h_{2}(O)=YL^\ast$, we have 
\begin{align*}
\psi^g &= c_{1}{E\{h_{1}(O)\}} +  c_{2} {E[E\{h_{2}(O)\mid A=a,\bL \} ]}
\nonumber
\\& = \underbrace{E\{Y(1-L^\ast)\}}_{\nu_{1}} + \underbrace{E[E\{YL^\ast\mid A=1,L\}]}_{\nu_{2}} 
\end{align*}
by Theorem \ref{theorem:thmone} and further, the EIF for $\psi^g$ is given by
\begin{align*}
U_{\psi^g}(O) = Y(1-L^\ast) + \frac{AL^\ast}{f(A\mid \bL)} \left\{Y - m(A,\bL)\right\} + m(1,\bL)L^\ast- \psi^g.
\end{align*}
This can be re-expressed as 
\[
U_{\psi^g}(O) = \frac{q^{g}(A\mid \textbf{L})}{f(A\mid \textbf{L})} \left\{Y - m(A,\textbf{L})\right\} + m(A,\textbf{L})(1-L^\ast) + m(1,\textbf{L})L^\ast - \psi^g
\]
which is a useful representation for deriving doubly robust estimators.
By Corollary \ref{corollary:corone}, estimators based on the EIF for $\psi^g$ in this case will be doubly robust because $\nu_{1}$ can be estimated non-parametrically, and EIF-based estimators for $\nu_{2}$ are doubly robust. In particular, these will be consistent if either $f(A\mid \bL)$ or $m(A,\bL)$ is consistently estimated, not necessarily both.

\begin{example}
Representative interventions for $J=1$. The intervention treatment distribution is given by $q^\text{g}(a\mid {\bl})=f(a\mid {\bl}, R=1)$ where $R=I(A\geq \delta)$.
\end{example}
In this case we have  
\begin{align*}
\psi^g &= E_{{\bL}}\left\{\sum\nolimits_{a=0}^1 E(Y\mid a,{\bL})q^g(a\mid {\bL})\right\} 
 = E_{{\bL}}\left[E_A\left\{ E(Y\mid A, {\bL}, R=1)\mid {\bL}, R=1 \right\}\right] 
= E_{{\bL}}\left\{ E(Y\mid {{\bL}}, R=1)\mid \right\}
\end{align*}
Following previous results \cite{Young2019}, we can see that, for this choice of intervention treatment distribution and $J=1$, $\psi^{g}$ is only a function of $R$, a coarsening of $A$, and takes the same form as the g-formula indexed by the static deterministic strategy that sets treatment to 1 but with $R$ playing the role of treatment.  
We can apply Theorem \ref{theorem:thmone} and Corollary \ref{corollary:corone} replacing the treatment $A$ with $R$ and $a^\ast$ with $r^\ast$.  Specifically, selecting  $r^\ast=1$, $c_{2}=1$, $h_{1}(O)=0$, $h_{2}(O)=Y$, by Theorem \ref{theorem:thmone} we have
\begin{align*}
\psi^g &= c_{2} {E[E\{h_{2}(O)\mid R=r^\ast,L \} ]} = \underbrace{E\left\{E(Y \mid R=1,\bL) \right\}}_{\nu_{2}}   
\end{align*}
and the EIF for $\psi^g$ is
\begin{equation*}
U_{\psi^g}(O)=\frac{I(R = 1)}{f(R\mid \bL)} \left\{Y - m(R,\bL)\right\} + m(1,\bL) - \psi^g
\end{equation*}
By Corollary \ref{corollary:corone}, estimators based on the EIF will be model doubly robust, i.e., consistent if either models for $f(R\mid L)$ \textit{or} $E(Y\mid R,L)$ are correctly specified.

\begin{example}
Multiplicative shift incremental propensity score interventions for $J=1$.
The intervention treatment distribution is given by $q^{g}(a\mid \bl)=(1-\delta)a l^\ast + f(a\mid \bl)(l^\ast \delta + 1 - l^\ast)$. Note the intervention distribution in Example 1 is a special case with $\delta=0$.
\end{example}

In this case, for a choice of $\delta\in\{0,1\}$ we have
\begin{align*}
\psi^g(\delta) &= E_{\bL}\left\{\sum\nolimits_{a=0}^1 E(Y\mid a,\bL)q^g(a\mid \bL)\right\} 
\\&= 
E_{\bL}\left[\sum\nolimits_{a=0}^1 E(Y\mid a,\bL)\left\{f(a\mid \bL)(L^\ast\delta+1-L^\ast)+L^\ast a(1-\delta)\right\}\right]\nonumber
\\& = E_{\bL,A}[E\left\{Y(L^\ast\delta+1-L^\ast)\right\}\mid A,\bL] + E_{\bL}\left[E\{YL^\ast(1-\delta)\mid A=1,L\} \right] \nonumber
\end{align*}
\sloppy
Selecting $a^\ast=1$, $c_{1}=1$, $c_{2}=(1-\delta)$, $h_{1}(O)=Y(L^\ast\delta+1-L^\ast)$, $h_{2}(O)=YL^\ast$, we have
\begin{align*}
\psi^g(\delta) &= c_{1}{E\{h_{1}(O)\}} +  c_{2} {E[E\{h_{2}(O)\mid A=a,L \} ]}
\nonumber
\\& = \underbrace{E\{Y(L^\ast\delta+1-L^\ast)\}}_{\nu_{1}} + (1-\delta)\underbrace{E[E\{YL^\ast\mid A=1,L\}]}_{\nu_{2}} 
\end{align*}
by Theorem \ref{theorem:thmone} and the EIF for $\psi^g(\delta)$ is given by
\begin{align*}
U_{\psi^g(\delta)}(O) = Y(L^\ast\delta+1-L^\ast) +  (1-\delta)\left[\frac{L^\ast A}{f(A\mid \bL)} \left\{Y - m(A,\bL)\right\} + m(1,\bL)L^\ast\right] - \psi^g(\delta).
\end{align*}
This can be re-expressed as
\begin{align*}
U_{\psi^g(\delta)}(O) =  \frac{q^g(A\mid \textbf{L})}{f(A\mid \textbf{L})} \left\{Y - m(A,\textbf{L})\right\} + m(A,\textbf{L}) (L^\ast\delta+1-L^\ast) + m(1,\textbf{L})L^\ast(1-\delta) - \psi^g(\delta)
\end{align*}
which is useful for deriving doubly robust estimators.
By Corollary \ref{corollary:corone}, the estimators based on the EIF for $\psi^g(\delta)$ will be model doubly robust.
As Kennedy (2019) \cite{Kennedy2019} noted, the EIF for $\psi^g$ indexed by an odds shift (\ref{kennedyshift}) is not model doubly robust and, therefore, does not meet the conditions of Corollary \ref{corollary:corone}.

Note that the conditions of Corollary \ref{corollary:corone} are sufficient for model double robustness of the EIF but are not necessary conditions. In Web Appendix B, we consider model double robustness and the EIF for deterministic strategies that depend on the natural value of treatment discussed in Section \ref{examples} \cite{Munoz2012,Haneuse2013,Young2014,Diaz2020}.
These are examples of $\psi^g$ indexed by intervention treatment distributions that do not meet the conditions of Corollary \ref{corollary:corone} yet model doubly robust estimators still exist.

\subsection{Time-varying treatments}
\label{sec:arg_long}

Recently, Molina (2017)\cite{Molina2017} showed that, in time-varying treatment settings, estimators derived from the EIF for a $\psi^{g}$ indexed by any intervention treatment distribution that does not depend on the observed treatment process \cite{Robins2000p,Bang2005,van2011} confer more protection against model misspecification than model double robustness. Rather, they showed that these estimators are $J+1$ model multiply robust, which implies model double robustness.
The following Theorem gives a sufficient condition for the existence of $J+1$ model multiple robust estimators of $\psi^{g}$ when the intervention treatment distribution may depend on the observed treatment process, and a simple approach to deriving the EIFs for a particular class of such intervention treatment distributions.

\begin{theorem}
\label{theorem: thmlong}
Suppose an intervention treatment distribution can be written as the following:
\begin{align}
q^{g}(a_j\mid 1, \bar{\bl}_{j},\bar{a}_{j-1}) 
=& c_1 h_1(\bar{\bl}_{j},\bar{a}_{j-1})I(a_j=a_j^\ast) + c_2 h_2(\bar{\bl}_{j},\bar{a}_{j-1})f(a_j\mid 1, \bar{\bl}_{j}, \bar{a}_{j-1}) + \label{eq:longtmt}
\\& c_{3} h_{3}(\bar{\bl}_{j},\bar{a}_{j-1})p^*(a_j\mid 1, \bar{\bl}_{j}, \bar{a}_{j-1})\nonumber
\end{align}
where $a_j^\ast$,
$c_1,~c_2$ and $c_{3}$ are constants; $h_1(\bar{\bL}_{j},\bar{A}_{j-1})$, $h_2(\bar{\bL}_{j},\bar{A}_{j-1})$ and $h_{3}(\bar{\bL}_{j},\bar{A}_{j-1})$ are known measurable functions of $(\bar{\bL}_{j},\bar{A}_{j-1})$; and $p^*(a_j\mid 1, \bar{\bl}_{j}, \bar{a}_{j-1})$ is a non-degenerate known probability distribution for $A_j$. Then the EIF for $\psi^{g}$ indexed by this intervention treatment distribution is 
\begin{equation}
\label{eq:genEIF}
\begin{aligned}
U_{\psi^g}(O) = 
\sum_{j=1}^{J}({T_j}-Q_{j-1})\prod_{k=0}^{j-1}\frac{q^{g}(A_k\mid Y_k=1, \bar{\bL}_k, \bar{A}_{k-1})}{f(A_k\mid Y_k=1, \bar{\bL}_k, \bar{A}_{k-1})} +
T_0 -\psi^g
\end{aligned} 
\end{equation}
where $Q_j\equiv Q_j(\bar{\bL}_j, \bar{A}_j, \bar{Y}_j)$ and $T_j\equiv T_j(\bar{\bL}_j, \bar{A}_j, \bar{Y}_j)$ are iteratively defined from $j=J-1,\ldots,0$ such that for $T_J \equiv Y_J$, we have $Q_j\equiv E(T_{j+1}\mid \bar{\bL}_j, \bar{A}_j, \bar{Y}_j)$ and 
\begin{align*}
T_j = &c_1Q_j^{A_j=a_j^\ast}h_{1}(\bar{\bL}_{j},\bar{A}_{j-1}) + c_2Q_jh_{2}(\bar{\bL}_{j},\bar{A}_{j-1}) + 
 c_{3}\left\{\sum_{a_j}p^*(a_j\mid 1, \bar{\bl}_{j}, \bar{A}_{j-1}) Q_j^{A_j=a_j}\right\}h_{3}(\bar{\bL}_{j},\bar{A}_{j-1})
\end{align*}
with $Q_j^{A_j=a^\ast_j}\equiv Q_j(\bar{\bL}_j, A_j=a_j^\ast,\bar{A}_{j-1}, \bar{Y}_j)$. Estimators based on this EIF are $J+1$ model multiply robust in that they are consistent if models for $Q_j$ are correctly specified for $j={k,\ldots, J-1}$ and the observed treatment models are correctly specified from $j=0, \ldots,k-1$ (for $k=0, \ldots, J$), where $j=s,s-1$ is $\emptyset$ $\forall s$.
\end{theorem}
{Theorem \ref{theorem: thmlong} makes the derivation of the EIF and the corresponding estimators far more straightforward and accessible when intervention distributions are in the form given by (\ref{eq:longtmt}).}
In Web Appendix D we prove that Expression (\ref{eq:genEIF}) is the EIF under a nonparametric model that imposes no restriction on the observed data  law for $\psi^g$ indexed by (\ref{eq:longtmt}). In Web Appendix E we prove that estimators based on this EIF are $J+1$ model multiply robust. Note that, by the monotonicity of the survival indicators, we have $Y_{j+1}=Y_{j+1}Y_j$. This implies that $Q_j = Y_j Q_j = Y_j Q_j(\bar{\bL}_j, \bar{A}_j, {Y}_j=1)$, where $Q_j(\bar{\bL}_j, \bar{A}_j, {Y}_j=1) = E(T_{j+1}\mid \bar{\bL}_j, \bar{A}_j, {Y}_j=1)$.
We now illustrate Theorem 2 in an example where the intervention treatment distribution depends on the observed treatment process. 
\begin{example}
Consider the multiplicative shift incremental propensity score interventions from Section \ref{examples}, recalling the intervention distribution is $q^{g}(a_j\mid 1, \bar{\bl}_j, \bar{a}_{j-1}) =  (1-\delta) l_j^\ast a_j  + (l_j^\ast \delta + 1 - l_j^\ast) f(a_j\mid 1, \bar{\bl}_{j},\bar{a}_{j-1})$. 
\end{example}

This intervention distribution can be written in the form of Equation (\ref{eq:longtmt}) by selecting $a_j^\ast=1$, $c_1=1-\delta,~c_2=1,~h_1(\bar{\bl}_{j},\bar{a}_{j-1})=l_j^\ast,~h_2(\bar{\bl}_{j},\bar{a}_{j-1})=l_j^\ast \delta+1-l_j^\ast,~h_3(\bar{\bl}_{j},\bar{a}_{j-1})=0$.
By Theorem \ref{theorem: thmlong}, the EIF for this intervention distribution is then given by:

\setlength{\abovedisplayskip}{1pt}
\begin{equation}
\label{eq:genMIEIF}
\begin{aligned}
U_{\psi^g(\delta)}(O) = &(Y_J - Q_{J-1}) \prod_{j=0}^{J-1}\frac{q^{g}(A_j\mid Y_j=1, \bar{\bL}_j, \bar{A}_{j-1})}{f(A_j\mid Y_j=1,  \bar{\bL}_j, \bar{A}_{j-1})} + 
\\&\sum_{j=1}^{J-1}\Big\{\underbrace{ (1-\delta) Q_j^{A_j=1}L_j^\ast + Q_j(L_j^\ast\delta+1-L_j^\ast) }_{T_j}-Q_{j-1}\Big\}\prod_{k=0}^{j-1}\frac{q^{g}(A_k\mid Y_k=1, \bar{\bL}_k, \bar{A}_{k-1})}{f(A_k\mid Y_k=1, \bar{\bL}_k, \bar{A}_{k-1})} +
\\&\underbrace{(1-\delta)Q_0^{A_0=1}L_0^\ast + Q_0(L_0^\ast\delta + 1-L_0^\ast)}_{T_0} -\psi^g
\end{aligned} 
\end{equation}
It is also straightforward to see that any $g$ corresponding to a deterministic static treatment rule meets the conditions of Theorem 2 by selecting $h_2(\bar{\bl}_{j},\bar{a}_{j-1})=h_3(\bar{\bl}_{j},\bar{a}_{j-1})=0$, $h_1(\bar{\bl}_{j},\bar{a}_{j-1})=1$ and $c_1=1$.  
In Web Appendix E, we further illustrate the application of Theorem \ref{theorem: thmlong} to deterministic dynamic treatment rules, as well  as other examples of intervention distributions that depend on the observed treatment process for the time-varying case including representative interventions and dynamic treatment initiation strategies with a grace period. 
Note that, in these examples and the incremental propensity score intervention example above, Theorem 2 holds by selecting $h_3(\bar{\bl}_{j},\bar{a}_{j-1})=0$ such that $p^*(a_j\mid 1, \bar{\bl}_{j}, \bar{a}_{j-1})$ need not be specified. More generally, the applicability of Theorem 2 may require specification of $p^*(a_j\mid 1, \bar{\bl}_{j}, \bar{a}_{j-1})$.  For example, this applies to an alternative grace period strategy where initiation within the grace period is assigned such that there is a uniform probability of initiating at each $j$ \cite{Cain2010}. 

Note that the EIF given in Diaz et al. (2020) \cite{Diaz2020} cover a comprehensive class of interventions that also guarantees estimators with model double robustness, including interventions that depend on the natural value of treatment \cite{Munoz2012,Haneuse2013} also known as modified treatment policies. 
Following the results of Theorem 2 in Diaz et al. (2020), the EIF of the implied modified treatment policy from our proposed intervention necessarily involve randomizer terms, but their derivation of the corresponding EIF assumes that the distributions of the randomizers are not known, when they certainly will be.
Our Theorem 2 provides the EIF for the g-formula indexed by a class of stochastic interventions that may depend on the observed treatment process. 
It can be shown that nearly all functionals in this class are captured by the g-formula functionals for which Diaz et al. (2020) provides the EIF.  Diaz et al.'s results would capture all of the functionals in this class, including those indexed by our proposed multiplicative incremental propensity score interventions, provided they projected their EIF onto a tangent space corresponding to smaller models whereby the distributions of some conceptualized randomizers are known. 
We do not take this approach and allow one to derive the EIF directly from a stochastic treatment distribution without requiring one to define an implied modified treatment policy first\footnote{There are multiple ways to define a modified treatment policy that has an identifying formula to equal the g-formula for our multiplicative shift intervention. It will involve randomizer terms that can potentially also depend on $L_j^*$.} This alternative derivation of the EIF may be more intuitive for treatment distributions that do not depend on the natural value of treatment.
    
Finally, in Web Appendix C we use a similar line of reasoning to Theorem \ref{theorem:thmone} and Corollary \ref{corollary:corone} to derive the EIF for $\psi^g$ and to assess the existence of doubly robust estimators for $\psi^{g}$ indexed by an intervention distribution that depends on the observed treatment process when $J=2$ with examples. However, this approach to deriving the EIF is cumbersome for large $J$, providing no simplification over Theorem \ref{theorem: thmlong}. 

\section{Estimators of $\psi^g$ indexed by multiplicative shift incremental propensity score interventions}
\label{sec:proposedint}
In this section, we consider various estimators of $\psi^g$ under the multiplicative shift incremental propensity score interventions defined by (\ref{ourshift}).  

\subsection{EIF-based estimators}
Several EIF-based estimators for $\psi^{g}$ have been proposed for deterministic treatment interventions including Bang and Robins (2005)'s estimator \cite{Bang2005,Scharfstein1999,Robins2000}, weighted ICE estimator \cite{Robins2007, Rotnitzky2017} and targeted maximum likelihood estimator (TMLE) \cite{van2011,Petersen2014,Lendle2017}. 
Weighted ICE and TMLE are variations of Bang and Robins (2005)\cite{Bang2005}. Compared with Bang and Robins (2005)\cite{Bang2005}, weighted ICE can give better performance \cite{Tran2019}.  Unlike both Bang and Robins (2005)\cite{Bang2005} and weighted ICE, TMLE can incorporate machine learning algorithms \cite{Tran2019}. In the absence of machine learning algorithms, weighted ICE and TMLE perform similarly \cite{Tran2019}, but weighted ICE is easier to implement.
In this section we will consider two estimators: 1) weighted ICE estimator that uses parametric models to estimate the nuisance functions thereby allowing for $J+1$ model multiple robustness; and 2) TMLE that also uses sample-splitting and cross fitting \cite{Van2000,Zheng2010,Chernozhukov2018} to allow one to incorporate machine learning algorithms to estimate the the nuisance functions.

\subsubsection{Weighted ICE estimator}
\label{sec:wice}
Let $\pi_j\equiv f(A_j\mid Y_j=1, \bar{A}_{j-1}, \bar{\bL}_j)$ and let $\pi_j(\alpha_j) = f(A_j\mid Y_j=1, \bar{A}_{j-1}, \bar{\bL}_j; \alpha_j)$ be a working parametric model for $\pi_j$ with $\alpha = (\alpha_0,\ldots,\alpha_{J-1})$. 
Denote estimates $\hat{\pi}_j\equiv \pi_j(\hat{\alpha}_j)$ of $\pi_j$ with $\hat{\alpha}_j$ the maximum likelihood estimate (MLE) of ${\alpha}_j$ computed from the observed data.
Subsequently, let $\hat{q}^{g}_j \equiv q^{g}_j(\hat{\pi}_j)$ be an estimate of $q^{g}(A_j\mid Y_j=1, \bar{A}_{j-1}, \bar{\bL}_j)$ as defined in (\ref{ourshift}) for a choice of $\delta\in[0,1]$, replacing the observed treatment process with the estimate $\hat{\pi}_j$.  
Let $\hat{Q}_j$ be a working parametric model for $Q_{j}$ defined in Theorem 2.  In the following algorithm, each $\hat{T}_j$ is calculated by replacing $Q_{j}$ in formula (\ref{eq:genMIEIF}) with the estimate $\hat{Q}_j$.  The weighted ICE algorithm is specifically implemented as follows:

\begin{algorithm}[H]                
\renewcommand{\theenumi}{\Alph{enumi}}   
\caption{Algorithm for Weighted ICE}          
\begin{algorithmic} [1]      
\item Compute the MLEs $\hat{\alpha}$ of ${\alpha}$ from the observed data. Set $\hat{T}_{J} = Y_J$. \\
Recursively from $j=J-1,\ldots,0$: 
\begin{enumerate}
    \item Fit a logistic regression model $Q_j(\bar{\bL}_{j},\bar{A}_{j}, Y_j=1;\theta_j)= \expit\{\theta_j^T\phi(\bar{\bL}_{j},\bar{A}_{j})\}$ for $E(\hat{T}_{j+1} \mid \bar{\bL}_{j},\bar{A}_{j}, Y_j=1)$ with observational weight $\prod_{k=0}^{j}({\hat{q}^{g}_k}/{\hat{\pi}_k})$ in those who survive by time $j$.
Here, $\phi(\bar{\bL}_j,\bar{A}_j)$ is a known function of $\bar{\bL}_{j}$ and $\bar{A}_{j}$.
More specifically, we solve for $\theta_j$ in the following estimating equation:
\begin{equation*}
   \mathbb{P}_n \left[ Y_{j}\prod_{k=0}^{j}\frac{\hat{q}^{g}}{\hat{\pi}_k} \phi_{j}({\bar{\bL}_{j},\bar{A}_{j}})\left\{\hat{T}_{j+1} - Q_{j}(\bar{\bL}_{j},\bar{A}_{j}, Y_j=1;{\theta}_{j})\right\}\right] = 0
\end{equation*}
\item Compute $\hat{T}_{j}$ from $\hat{Q}_{j}\equiv Q_{j}(\bar{\bL}_{j},\bar{A}_{j}, \bar{Y}_j;\hat{\theta}_{j})$ ensuring $\hat{T}_{j}=0$ when $Y_j=0$.
\end{enumerate}
\item Estimate $\hat{\psi}^g(\delta)_{WICE}=\mathbb{P}_n(\hat{T}_0)$
\end{algorithmic}
\end{algorithm}
\noindent where $\mathbb{P}_n\{f(X)\} =n^{-1}\sum_{i=1}^n f(X_i)$. Following arguments in Section \ref{sec:arg_long}, this estimator is $J+1$ model multiply robust.

\subsubsection{TMLE with sample-splitting and cross-fitting}
This algorithm utilizes sample-splitting and cross-fitting to allow flexible machine learning algorithms for estimating nuisance functions while circumventing Donsker class conditions \cite{Van1996,Van2000}. 
In Web Appendix G, we prove the asymptotic normality of this estimator under the condition that the nuisance functions are estimated consistently at rates faster than $n^{-1/4}$ when $\psi^{g}$ is indexed by the interventions (\ref{ourshift}). 

Suppose that a sample of size $n$ is split into $M$ disjoint subsets. Let $S_m$ denote the subset of individuals in split $m=1,\ldots,M$ and let $S_{-m}$ denote individuals not in split $m$ (i.e., $S_{-m} = \{i \notin S_m\}$).
Moreover, let $\hat{\pi}_j^{(-m)}$, $\hat{q}_j^{(-m)}$ and $\hat{Q}_j^{(-m)}$ denote estimates of $\pi_j,~q_j^g$ and $Q_j$ \textit{obtained from machine learning algorithms to individuals in $S_{(-m)}$}. 

\begin{algorithm}[H]  
\renewcommand{\theenumi}{\Alph{enumi}}                 
\caption{Algorithm for TMLE with sample-splitting and cross-fitting}
\begin{algorithmic}[1]      
\item For each $m=1,\ldots,M$:
\begin{enumerate}
    \item \textbf{\textit{For individuals in $S_{-m}$}}: compute $\hat{\pi}_j^{(-m)}$, $\forall j$. Set $\hat{T}_{J} = Y_J$. 
    \item Recursively from $j=J-1,\ldots,0$ \textbf{\textit{for individuals in $S_{-m}$}}: 
    \begin{enumerate}
        \item Compute $\hat{Q}_j^{(-m)}(\bar{\bL}_{j},\bar{A}_{j}, Y_j=1)$ by regressing $\hat{T}_{j+1}$ on $(\bar{\bL}_{j},\bar{A}_{j})$ in those alive at time $j$
        \item Compute $\hat{T}_{j}$ from $\hat{Q}_{j}^{(-m)} \equiv \hat{Q}_{j}^{(-m)}(\bar{\bL}_{j},\bar{A}_{j}, \bar{Y}_j)$ by formula (\ref{eq:genMIEIF}), setting $\hat{T}_{j}=0$ if ${Y}_{j}=0$
    \end{enumerate}
        \item \textbf{\textit{For individuals in $S_{m}$}}, set $\hat{T}_{J} = Y_J$. Then recursively from $j=J-1,\ldots,0$: 
    \begin{enumerate}
        \item Solve for $\gamma_j$ in the following set of estimating equations:
        \begin{equation*}
         \mathbb{P}_n^{m} \left( Y_{j}\prod_{k=0}^{j}\frac{\hat{q}_k^{{g}^{(-m)}}}{\hat{\pi}_k^{(-m)}} \left[\hat{T}_{j+1} -  \expit\left\{\logit\left(\hat{Q}_j^{(-m)}(\bar{\bL}_{j},\bar{A}_{j}, Y_j=1)\right)+\gamma_j \right\}\right]\right) = 0
        \end{equation*}
        \item Compute $\hat{T}_{j}$ from $\hat{Q}_j^\Delta(\bar{\bL}_{j},\bar{A}_{j}, Y_j=1) \equiv \expit\left\{\logit\left(\hat{Q}_j^{(-m)}(\bar{\bL}_{j},\bar{A}_{j}, Y_j=1)\right)+\hat{\gamma}_j \right\}$ if $Y_j=1$, otherwise set $\hat{T}_{j}=0$ if ${Y}_{j}=0$
    \end{enumerate}
\end{enumerate}
\item Calculate \[\hat{\psi}^g(\delta)_{TMLE} = \frac{1}{M}\sum_{m=1}^M\mathbb{P}_n^m(\hat{T}_0)\]
\end{algorithmic}
\end{algorithm}
\noindent Here $\mathbb{P}_n^{m}\{f(X)\} = \frac{1}{\mid S_m\mid}\sum_{i\in S_m} f(X_i)$ where ${\mid S_m\mid}=n/M$ is the cardinality of $S_m$.

\subsection{Singly robust estimators}
 We also consider less optimal but computationally simple singly robust estimators of $\psi^{g}$ indexed by (\ref{ourshift}).  An IPW estimator $\hat{\psi^g}_{IPW}(\delta)$ can be obtained by the product $\hat{\psi}^g_{IPW}(\delta) = \prod_{j=0}^{J-1} \hat{\Upsilon}^g_{IPW,j}(\delta)$ where $\hat{\Upsilon}^g_{IPW,j}(\delta)$ can be interpreted as an estimate of the discrete hazard at $j$ under a stochastic strategy $g$ where treatment assignment is a draw from (\ref{ourshift}) given the identifying conditions of Section \ref{sec:g-formula}.  Each $\hat{\Upsilon}^g_{IPW,j}(\delta)$ can be obtained by solving for $\Upsilon^g_{IPW,j}(\delta)$ in the following estimating equations:
\[
  \mathbb{P}_n \left[ Y_{j}\prod_{k=0}^{j}\frac{\hat{q}^g_k}{\hat{\pi}_k  } \{Y_{j+1}-\Upsilon^g_{IPW,j}(\delta)\} \right] =0
\]

Alternatively, the singly-robust ICE estimator, which we will denote $\hat{\psi}^g_{ICE}(\delta)$, can be obtained as a special case of the algorithm for weighted ICE above where the observational weights are set to 1.

\subsection{Censoring}\label{sec:censoring} Straightforward extensions of the identification arguments in Section \ref{sec:g-formula} in studies with censoring follow by implicitly including in $g$ a hypothetical intervention that eliminates censoring throughout follow-up \cite{Young2011} with straightforward extensions of the g-formula $\psi^{g}$, properties of its EIF and associated estimation procedures.  Briefly, denote $C_j$ as the indicator of censoring  by time $j$ and adopt the order $(L_j,A_j,C_{j+1},Y_{j+1})$.  Extensions to accommodate censoring for singly robust weighted estimators and the various EIF-based estimators considered, require, in addition to estimating $\alpha_j$ in $f(A_j\mid Y_j=1, \bar{A}_{j-1}, \bar{\bL}_j, C_j=0; \alpha_j)$, also estimating $\alpha^c_j$ in $P(C_{j+1}=1\mid C_{j}=0,\bar{A}_{j}, \bar{\bL}_{j}, Y_{j}=1; \alpha^c_{j})$ for $j=0,\ldots, J-1$ with $\alpha^c = (\alpha^c_1,\ldots,\alpha^c_{J})$. Further details of modifications to the weighted ICE and TMLE to accommodate censoring are provided in Web Appendix F.

\section{Simulation studies}
\label{sec:sim}
We conducted two different simulation studies. The first simulation study aims to compare the performance of the weighted ICE, IPW and ICE estimators when the nuisance functions are estimated through parametric models under various model misspecification scenarios. The second simulation study aims to compare the performance of TMLE with sample splitting and cross fitting, IPW and ICE when the nuisance functions are estimated through machine learning algorithms.

\subsection{Simulation study 1: using parametric models}
In this simulation study we compare the performance of the weighted ICE estimator with the singly robust estimators (IPW and ICE estimators) for $\psi^{g}$ indexed by the intervention distribution (\ref{ourshift}) which, under identifying conditions discussed in Section \ref{sec:g-formula}, equals the cumulative probability of survival at $J$ under an intervention that increases the probability of treatment initiation in those with $L_j^\ast=1$ as a function $\delta$.  Recall that this increase is defined such that decreasing values of $\delta$ correspond to an increasing probability of treatment initiation (with $\delta=1$ coinciding with no treatment intervention).

We simulated 1000 samples of $n=(500,~1000,~2500)$ individuals selecting $J=5$ and $\delta=(0.75,~0.50,~0.25)$.  We simulated the following variables: $(\bL_0,A_0,C_1,Y_1,\bL_1,A_1,...,C_{5},Y_{5})$, where $\bL_j = (L_j^\ast, L_{1j}, L_{2j})$ is the vector of measured confounders. 
Specifically, we generated $L_{1}^\ast$ and $L_{11}\sim \text{Ber}\{\text{expit}(-1)\}$, and $L_{2}\sim \text{Ber}\{\text{expit}(1+L_{1}^\ast)\}$. 
The censoring indicator at each time $j$ ($j=1,\ldots,5$) was simulated from $C_j \sim \text{Ber}\{\text{expit}(-2+L_{1j}-L_{2j})\}$ if $C_{j-1}=0$ and $Y_j=1$. 
The outcome at each time $j$ ($j=1,\ldots,5$) is simulated from $Y_j \sim \text{Ber}\{\text{expit}(1+3A_{j-1}-2L_{j-1}^\ast+L_{1,j-1}-L_{2,j-1})\}$ if $Y_{j-1}=1$ and $C_{j}=0$. 
The time-varying confounders at time $j$ ($j=0,\ldots,4$) are simulated from $L_{j}^\ast \sim  \text{Ber}\{\text{expit}(-1-A_{j-1}+L_{j-1}-L_{1,j-1}+L_{2,j-1})\}$, $L_{1j} \sim  \text{Ber}\{\text{expit}(-1+A_{j-1}+L_{1,j-1}-L_{2,j-1})\}$ and $L_{2j} \sim  \text{Ber}\{\text{expit}(1+A_{j-1}+L_j^\ast+L_{2,j-1})\}$ if $Y_j=1$.
Treatment at time $j$ ($j=0,\ldots,4$) is simulated from $A_j \sim \text{Ber}\{\text{expit}(-1-2L_j^\ast-L_{1j} + L_{2j} + 2A_{j-1})\}$ if $Y_j=1$.
In addition $(Y_j, \bL_{j}, A_j,\ldots)=(\emptyset,\emptyset,\emptyset,\ldots)$ if $C_j=1$, and $(\bL_{j}, A_j, C_{j+1},\ldots)=(\emptyset,\emptyset,\emptyset,\ldots)$ if $Y_{j}=0$.

\sloppy
The true cumulative probabilities of survival were calculated by using the true parametric models to generate a Monte Carlo sample of size $10^7$ under all interventions of interest. Our selection of parameters resulted in a scenario where selecting smaller $\delta$ (that is, interventions with larger increases in the probability of treatment initiation at each $j$) improves survival. 

We considered three estimation scenarios for each choice of $\delta$ and sample size such that (1) all models are correctly specified, (2) only the outcome regression  models are correctly specified, and (3) only the treatment (propensity score) and censoring models are correctly specified.  The true functional forms of the treatment and censoring models are known under our simulation because treatment and censoring were generated to only depend on past measured variables. Similarly, the functional form of the outcome regression model for $Q_{J-1} = E(Y_J \mid Y_{J-1}=1, C_J=0, \bar{A}_{J-1}, \overline{L}_{J-1})$ is known due to the absence of unmeasured common causes. However, the true functional forms of the outcome regressions $Q_j$ for $0\leq  j<J-1$ are not known under our simulation.  To ensure correctly specified models for $Q_j$, $0\leq  j<J-1$, saturated models were fit, i.e., all main terms and interaction terms for $(A_{j},L_{j}^\ast,L_{1j},L_{2j})$.  In scenarios with misspecified models, at each time $j$, the misspecified treatment model ignores the censoring process and excludes $A_{j-1}$ in the model, and the misspecified outcome regression model excludes any pairwise interactions between the covariates and treatment. 
\begin{figure}
\includegraphics[width=15cm]{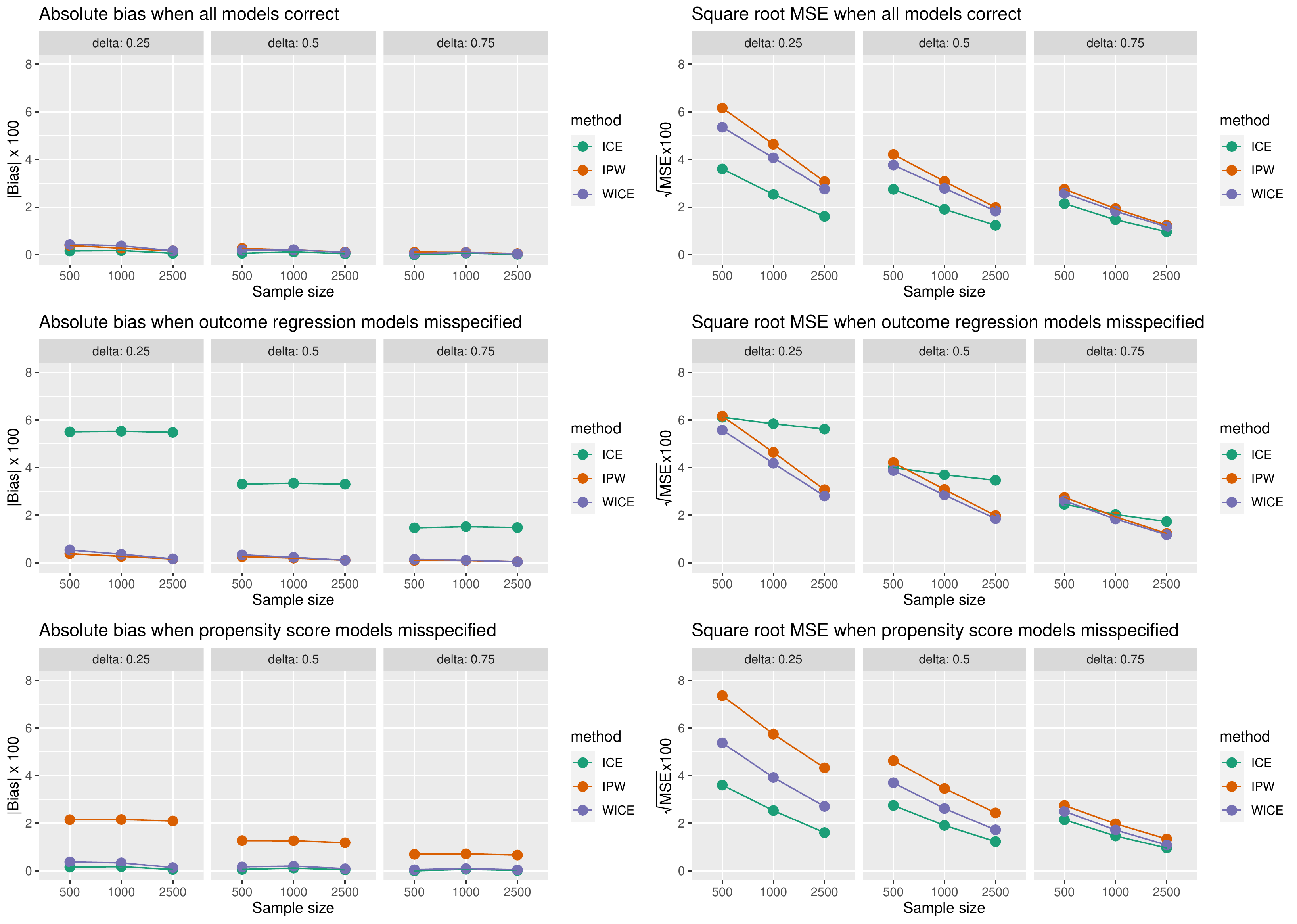}
\caption{Results for the Simulation study 1}
\label{fig:paramsim}
\end{figure}

Figure \ref{fig:paramsim} compares performance of the three estimators of  $\psi^{g}$ indexed by (\ref{ourshift}) for $\delta=(0.75,~0.50,~0.25)$. 
Complementary results are given in Tables 3--5 in Web Appendix H. 
As expected, all estimators were nearly unbiased under correctly specified models. 
Under our model misspecification scenarios, $\hat{\psi}^g(\delta)_{WICE}$ is nearly unbiased, but the IPW estimator is biased when the treatment models are misspecified, and the ICE estimator is biased when the outcome models are misspecified.
In addition, under correctly specified models $\hat{\psi}^g(\delta)_{ICE}$ is the most efficient, and $\hat{\psi}^g(\delta)_{IPW}$ is the least efficient estimator. Interestingly, the simulation results show that $\hat{\psi}^g(\delta)_{WICE}$ has smaller MSE than the IPW estimator in all scenarios.

The simulation results also show that as $\delta$ decreases, the standard error (and MSE) in all three estimators increases. This is due to an increase in the effect of near positivity violations as $\delta$ nears zero. In fact, we would expect all three estimators to have the largest standard errors when $\delta=0$, which is equivalent to a strategy that treats all individuals with $L_j^\ast=1$ at all times.  In Web Appendix H, we also show $J+1$ model robustness of our weighted ICE estimator in a model misspecification scenario that requires more than model double robustness.

\subsection{Simulation study 2: using machine learning methods}
In the second simulation, we compare the performance of algorithms that use machine learning to estimate the nuisance functions for $\psi^{g}$ indexed by (\ref{ourshift}) with $J=5$. Specifically, we compare TMLE with sample splitting and cross fitting, IPW and ICE. Given much longer computation times, we limited consideration to one choice of $\delta=0.5$.  
Unlike Simulation 1, we add model complexity to the data generating mechanism by considering continuous covariates, which might mimic real-life data more closely. We simulated 1000 hypothetical cohorts of $n=(250,~500,~1000)$ comprising the following variables: $(\bL_0,A_0,C_1,Y_1,\bL_1,A_1,...,C_{5},Y_{5})$, where $\bL_j = (L_{1j}, L_j^\ast)$. In addition, $\bL_0=(L_0^{1}, L_0^{2}, L_{10}, L_0^\ast)$, where $L_0^{1}$ and $L_0^{2}$ are baseline covariates. 
In particular, $L_0^{1}\sim \text{Ber}(0.5)$, $L_0^{2}\sim \text{N}(0,1)$, $L_{10}\sim \text{N}(2+L_0^{1},1)$ and $L_0^\ast\sim \text{Ber}\{\text{expit}(1.5-0.5L_0+L_0^{1}+0.25L_0^{2})\}$. 
For $j\geq 1$, 
$L_{1j}\sim \text{N}(2+A_{j-1}-L_{j-1}^\ast+0.5L_{1,j-1}+L_0^{1},1)$ and $L_{j}^\ast \sim  \text{Ber}\{\text{expit}(1.5-A_{j-1}-0.5L_{1j}+L_{j-1}^\ast+L_0^{1}+0.25L_0^{2})\}$ if $Y_j=1$.
Censoring indicator at each time $j$ ($j=1,\ldots,5$) is simulated from $C_j \sim \text{Ber}[\text{expit}\{-4-A_{j-1}-L_{j-1}^\ast-0.5\sqrt{\mid L_{1,j-1}L_0^{2}\mid} + 1.5\mid L_{1,j-1} \mid/(1+\exp(L_0^{2}))\}]$ if $C_{j-1}=0$ and $Y_j=1$. 
The outcome at each time $j$ ($j=1,\ldots,5$) is simulated from $Y_j \sim \text{Ber}[\text{expit}\{-1+2A_{j-1}-2L_{j-1}^\ast+0.25L_{j-1}^\ast L_{1,j-1}+0.5L_0^{1}+0.75\mid L_{1,j-1}+L_0^{2}\mid^{1.5}\}]$ if $Y_{j-1}=1$ and $C_{j}=0$. 
Treatment at time $j$ ($j=0,\ldots,4$) is simulated from $A_j \sim \text{Ber}\{\text{expit}(-3+L_{j}^\ast-0.5L_{1j}+0.25L_{j}^\ast L_{j}+0.5L_0^{1}+0.25L_0^{2} + 0.5\mid L_0^{2}\mid)\}$ if $Y_j=1$ and $A_{j-1}=0$, and is set to $1$ if $Y_j=1$ and $A_{j-1}=1$.

Nuisance functions were estimated using the Super Learner ensemble, which uses cross validation to select the best convex combination of predictions from a pool of prediction algorithms \cite{van2007}. The library of potential candidates used here consisted of: generalized linear models and its variants (SL.glm, SL.glm.interaction), Bayesian generalized linear models (SL.bayesglm), generalized additive models with smoothing splines (SL.gam), multivariate adaptive regression Splines (SL.earth), neural networks (SL.nnet) and random forest (SL.ranger). 

Table~\ref{tab:ML} compares the performance of the 3 estimators.  The ICE and IPW estimators show bias as they are not expected to converge at $\sqrt{n}$ rates when machine learning is used for nuisance parameter estimation.  TMLE, on the other hand, show little to no bias in all instances. This agrees with theory as TMLE allows the nuisance functions to converge at slower nonparametric rates.
Moreover, the estimated coverage probability of the confidence intervals for TMLE based on the asymptotic variance (see Web Appendix G) is very close to the nominal 95\%: $(94.7,~96.2,~95.2,~94.4)$ for $n=(250,~500,~1000,~2500)$, respectively.

\setlength{\tabcolsep}{1.5pt}
\renewcommand{\arraystretch}{1.0}
\begin{table}
 \caption{\label{tab:ML}Simulation study 2 for proposed treatment intervention distribution and incorporating machine learning algorithms ($M=2$).
 True probability of survival at time 5 is $0.629$. All values are multiplied by 100.}
\centering{\scalebox{1}{
\begin{tabular}{| l | lcc  | lcc | lcc | lcc |}
\hline
& \multicolumn{3}{c}{$n=250$} & \multicolumn{3}{|c|}{$n=500$}  & \multicolumn{3}{|c|}{$n=1000$} & \multicolumn{3}{|c|}{$n=2500$}
\\ \hline
Estimator & BIAS & SE & RMSE & BIAS & SE & RMSE & BIAS & SE & RMSE & BIAS & SE & RMSE \\ 
\hline
$\hat{\psi}^g(\delta)_{ICE}$	&-1.50	&4.35	&4.61	&-0.82	&2.83	&2.95	&-0.47	&2.14	&2.19	&-0.16	&1.38	&1.39\\
$\hat{\psi}^g(\delta)_{IPW}$	&-1.50	&4.91	&5.13	&-1.50	&3.32	&3.64	&-1.35	&2.60	&2.93	&-1.10	&1.71	&2.03\\
$\hat{\psi}^g(\delta)_{TMLE}$	&-0.19	&5.79	&5.79	&-0.09	&3.61	&3.62	&-0.07	&2.59	&2.59	&0.03	&1.65	&1.65\\
\hline
\end{tabular} }}
\end{table}

\section{Application}
\label{sec:app}
Randomized trials suggest that antiretroviral pre-exposure prophylaxis (PrEP) is highly effective in preventing HIV infection among men who have sex with men (MSM) \cite{Grant2010,Molina2015,Mccormack2016}. At the same time, there is widespread concern that PrEP may decrease condom use,
thereby increasing incidence of bacterial sexually transmitted infections (STIs) among MSM \cite{Tellalian2013,Krakower2014}.
With this backdrop, PrEP uptake has been low in practice, particularly in MSM with markers for higher HIV risk \cite{Marcus2016}.  This is precisely the setting where near or true positivity violations will occur if the analyst attempts to query observational data about the effects of deterministic interventions such as ``always treat'' versus ``never treat'' with PrEP because propensity scores will be close to (or equal to) zero for individuals with certain levels of the measured confounders.  In conjunction with these challenges, such deterministic treatment effects are not of greatest interest for treatments like PrEP, where biological benefits are established but population disease burden may be impacted with even small increases in treatment uptake.

We illustrate the application of the estimators of survival by $J$ discussed in Section \ref{sec:proposedint} using electronic health record data from the Cambridge Health Alliance -- a large community healthcare system in Eastern Massachusetts -- to estimate effects of  increasing PrEP uptake on bacterial STI diagnosis. Specifically, we consider interventions that, beginning at the time of an HIV negative test, successfully increase the proportion initiating PrEP ($A_j$) in each follow-up week $j$ only among those receiving an STI \textsl{test} and no prior diagnosis of HIV at time $j$ ($L_j^{\ast}=1$, being tested for STIs suggests recent condomless sex and PrEP would not be used after an HIV diagnosis). No intervention is made for the remainder of the population at time $j$ ($L_j^{\ast}=0$).  Increases in treatment uptake under these interventions are quantified by a specified $\delta\in[0,1]$ as defined in (\ref{sec:proposedint}), which quantifies the factor by which the probability of treatment non-initiation is decreased (relative to no intervention) at $j$.   We consider $J=26$ weeks and, as in simulation study 1, consider $\delta=(0.95,~0.85,~0.75)$ representing realistic interventions that result in ``low'', ``medium'' and ``high'' success in PrEP uptake relative to no intervention.  We use $\delta=1$ (corresponding to no intervention) as the reference in defining causal effects.  

Our analytic data set was restricted to patients who met all the following inclusion criteria at some point during 2012--2017: 1) Cis male with report of male gender of sex partner(s); 2) 15 years of age or older; 3) an HIV-negative test; 4) had no PrEP prescription in the 3 months prior to baseline; and 5) had no STI diagnosis in the 12 months prior to baseline.
Baseline (week $j=0$) for an individual was defined as the first week that all of these inclusion criteria are met. 
For simplicity, we excluded one individual who met these criteria but died without a bacterial STI diagnosis during the 26 week follow-up period.  
Our final analytic data set consisted of $n=1103$ individuals.  As expected, few initiated PrEP over the follow-up (cumulatively 5.1\% over the 26 weeks). 
The cumulative proportion of those receiving an STI test while being HIV-free over the 26 weeks was 70.7\%.
Note that no individual was treated as censored in this analysis, requiring additional assumptions that medical care was not sought outside of the Cambridge Health Alliance by any individual included at baseline over the 26 week follow-up.  

Baseline covariates $\bL_0$ included age and calendar year at baseline, race/ethnicity, and time-varying covariates $\bL_j$ included indicator of any ambulatory encounter, indicator of HIV, indicator of any HIV testing and indicator of any STI testing. 
We used the Super Learner ensemble (with the same potential candidates as in the simulation) to estimate all nuisance functions for TMLE with $M=5$. We compared these results with the IPW, ICE and weighted ICE estimators described in this paper where the nuisance functions are specified by parametric models.
Confidence intervals for each of the methods are obtained from 1000 bootstrap samples by taking the 2.5th and 97.5th percentiles of the resulting estimates.

Our estimate of the probability of not receiving an STI diagnosis under no intervention by 26 week follow-up ($\delta=1$)  was 93.7\%.  Table \ref{tab:MLdata} shows results from the four methods for $\delta<1$.
In this case point estimates from all of the methods are similar.
The results do not provide sufficient evidence that increasing PrEP uptake increases risk of STI diagnosis. For instance, compared with no intervention ($\delta=1$), the relative survival estimates under low, medium and high increases in PrEP uptake were 0.99 (95\% CI = $(0.96,~1.01)$), 0.97 (95\% CI = $(0.91,~1.01)$) and 0.96 (95\% CI = $(0.87,~1.02)$), respectively under $\hat{\psi}^g(\delta)_{TMLE}$. The relative survival estimates using other estimators were very similar (see Web Appendix I). 
We also note that due to an increase in the presence of near positivity violations as $\delta$ nears zero, observational weights calculated under smaller $\delta$ were more variable than larger $\delta$ (see Web Appendix I). We would expect standard errors from all of the estimators to be the largest for $\delta=0$.

\setlength{\tabcolsep}{1.5pt}
\renewcommand{\arraystretch}{1.0}
\begin{table}
 \caption{\label{tab:MLdata}Point estimates and 95\% confidence intervals from analysis of MSM from the Cambridge Health Alliance on the effect of incremental PrEP initiation on incident STI diagnosis. All values are multiplied by 100. 
 }
\centering{\scalebox{1}{
\begin{tabular}{| l | cc  | cc | cc | cc |}
\hline
& \multicolumn{2}{c}{$\hat{\psi}^g(\delta)_{TMLE}$ (with ML)} & \multicolumn{2}{|c|}{$\hat{\psi}^g(\delta)_{WICE}$}  & \multicolumn{2}{|c|}{$\hat{\psi}^g(\delta)_{ICE}$} & \multicolumn{2}{|c|}{$\hat{\psi}^g(\delta)_{IPW}$}
\\ \hline
$\uparrow$ in PrEP & Est. & 95\% C.I. & Est. & 95\% C.I. & Est. & 95\% C.I. & Est. & 95\% C.I. \\ \hline
Low &92.9&(89.3,~95.2)&93.0&(90.9,~94.9)&92.9&(91.1,~94.6)&92.9&(90.5,~94.9)\\
Medium &91.4&(84.8,~95.5)&91.6&(87.1,~95,1)&91.4&(88.5,~94.1)&91.3&(85.8,~95.0)\\
High &90.9&(80.9,~95.8)&90.3&(83.2,~95.4)&90.0&(85.8,~93.8)&89.9&(80.9,~95.3)\\
\hline
\end{tabular} }}
\end{table}

\section{Discussion}
\label{sec:discussion}

Many methods that have been proposed for estimating causal estimands in time-varying treatment settings for survival analysis, and among these methods are estimators that offer protection against model misspecification and can also attain semiparametric efficiency bound. However, most of these doubly robust estimators have been in the setting of deterministic treatment interventions. In this paper we provided some sufficient conditions for the existence of doubly robust estimators when a treatment intervention distribution can depend on the observed treatment process for point treatment processes. 
We also discussed a class of intervention distributions that are always guaranteed to give doubly/multiply robust estimators and gave a general form of the EIFs that are associated with these intervention distributions.
Among these intervention distributions is our multiplicative shift incremental propensity score intervention distribution, which aims to increase treatment uptake in a group of individuals who are at high risk of the outcome but have low exposure to treatment.
We provided various estimators that can be used for our proposed treatment intervention for both parametric and machine learning algorithms.

We conducted two simulation studies for our proposed multiplicative shift intervention distribution. 
Our first study show that the weighted ICE is more robust to model misspecification than IPW and ICE when the nuisance functions are estimated using parametric models. 
Our second study show that the TMLE with sample-splitting and cross-fitting is consistent as long as the nuisance functions are estimated consistently at fast enough rates using machine learning methods, which may not necessarily be $n^{-1/2}$.
We also illustrated our an application of our estimators to a real world dataset in the PrEP context.

Note that our proposed intervention treatment distribution (\ref{ourshift}) is guaranteed under a stochastic intervention such that treatment initiation status at time $j$ for an individual with $L^{\ast}_j=1$ is a random draw from (\ref{ourshift}). The identifying conditions reviewed in Section \ref{sec:g-formula} are sufficient to give our effect estimates this interpretation.  A more policy relevant interpretation might, for example, be an intervention where individuals with $L^{\ast}_j=1$ are always offered PrEP counseling.  In these individuals, the intervention distribution (\ref{ourshift}) quantifies the hypothesized ``success'' of such an intervention where $g$ in this case really refers to a deterministic strategy relative to the unmeasured treatment ``offered PrEP counseling''.  Additional assumptions are needed to give our effect estimates this interpretation following similar arguments to those given in Richardson and Robins (2013)\cite{Richardson2013} and Young et al. (2014) \cite{Young2014}.

Finally, we note that while machine learning algorithms are more robust to model form misspecification, they are also computationally complex and may be practically infeasible for very large datasets without powerful computing systems. 
Moreover, issues related to data privacy make access to advanced computational resources impossible in many cases. Therefore, estimators that offer model double/multiple robustness are useful in practice as they offer protection against model misspecification and can be easily computed using standard off-the-shelf regression software in R.

\section*{Acknowledgment}
This research was funded by NIAID grant R21AI143386-01A1. The authors thank Dr. Gerard Coste at Cambridge Health Alliance for assistance with collection of clinical data. 


\bibliographystyle{SageV}
\bibliography{refs}

\end{document}